\documentstyle[epsf,twoside,fleqn,espcrc2]{article}

\newcommand{\AmS}{{\protect\the\textfont2
  A\kern-.1667em\lower.5ex\hbox{M}\kern-.125emS}}

\title{The Renormalization Group and Dynamical Triangulations}

\author{R. L. Renken\address{Department of Physics, University of Central 
Florida, Orlando, FL  32816}
	}
       
\begin{document}

\begin{abstract}
A block spin renormalization group approach is introduced which can be
applied to dynamical triangulations in any dimension.
\end{abstract}

\maketitle

\section{INTRODUCTION}
Given a general action for a theory of dynamical triangulations,
two limits must be taken to obtain a physical theory.  The cosmological
constant must be adjusted
to give the infinite volume limit (or some finite target volume in practical
calculations).  The remaining parameters in the action must be used to find a
second order phase transition so that a continuum limit of the lattice theory
can be taken.  The renormalization group approach is a natural technique
with which to search for and study the required critical phenomena [1-4].

The renormalization group approach can be viewed as a black box that takes
some initial theory $S$ with correlation length $\xi$ and produces an effective
theory $S^{\prime}$ with correlation length $\xi^{\prime} = \xi/b$.  $b$ is
fixed and is a property of the chosen renormalization group transformation.
By taking the output of the block box and feeding it back into the input, it
is possible to produce a sequence of theories: $S^{(0)}$,
$S^{(1)}$, $S^{(2)},\cdots$ with a corresponding sequence of correlation
lengths $\xi^{(0)}$, $\xi^{(1)}$, $\xi^{(2)},\cdots$.
If the original theory has
a finite correlation length, the $\xi^{(n)} \rightarrow 0$ as the
renormalization group transformation is iterated.  Such a theory is referred
to as a trivial theory.  If the original theory has an infinite correlation
length, as occurs at a second order phase transition, then the correlation
length stays infinite after each renormalization group transformation and 
what happens is that the effective theory approaches a fixed point,
$S^{(n)} \rightarrow S^*$.

In an ordinary statistical mechanical model, such as the Ising model, the
degrees of freedom can be organized into fixed blocks and then averaged
according to some rule to produce a block degree of freedom.  If the blocks
are, for instance, of volume $b^D$, then the scale factor is $b$ as above.
Dynamical triangulations are different in that there is apparently no way
to draw fixed boxes around the degrees of freedom, which are determined by
the connectivity of the lattice.  A related difficulty is that the Hausdorff
dimension is not known a priori so that rescaling a volume by some factor does
not determine what the length is rescaled by.  A related difficulty is
that the number of configurations cannot be enumerated in the trivial way they
can for ordinary statistical mechanical models.  In fact, for three and four
dimensions, it is not even known with certainty that this number grows only
exponentially with the volume.

A dynamical triangulation is interpreted as a lattice representation of a
spacetime with a metric.  Nodes connected by a link are considered closer than
those that are not connected by a link.  One way to define a renormalization
group transformation for dynamical triangulations is to insist that the block
triangulation preserve this physical notion.  Such a transformation succeeds
at getting critical couplings in two dimensions, but exponent calculations
fail to converge and the method is difficult to generalize to arbitrary $D$
[5,6].

Another renormalization group transformation in two dimensions is based on
the idea of removing a node and its associated triangles from the lattice
manifold and filling the hole back in with triangles but without any new
nodes [7].  This is achieved by making flips of links connected to the node in
question until its coordination number is three.  It and its three associated
triangles are then removed and replaced with a single triangle.  This method
correctly produces both critical couplings and critical exponents.

It is possible to make this approach work in arbitrary dimensions [8].  In
general $D$, the possible update moves can be labeled by $i$ with
$i=0,1,\cdots,D$.  An update move labeled by $i$ swaps an $i$-dimensional
subsimplex of a randomly chosen simplex with a $(D-i)$-dimensional subsimplex.
This eliminates $D-2i$ simplexes.  In two dimensions, a link flip is an
$i=1$ move while node deletion is an $i=0$ move.  In a two-dimensional ensemble
with fixed volume these are the only choices.  In order to eliminate a node
in general dimensions, some algorithm for choosing $i$ is necessary since there
are more possible values.  It is best to use moves with the lowest value
of $i$ possible, since these remove the most simplexes. $i=0$ is only possible
on the last move (when the node has coordination number $D+1$) and $i=1$ is
not always possible due to geometrical constraints and the existence of
connections on the surface around the node.  In fact, it is sometimes
necessary to make an $i=D$ move, which is node insertion.

\begin{figure}
\centering
\epsfxsize=3.0in \epsfbox{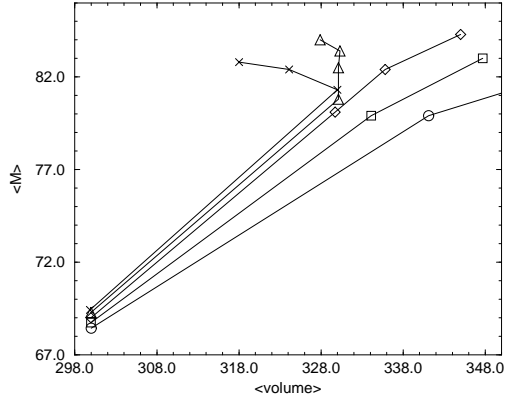}
\caption{Renormalization group flows in the four-dimensional dynamical
         triangulation model near the transition.}
\label{fourd}
\end{figure}

While deletion of a single node is viewed as a renormalization group
transformation resulting in a block lattice, it is generally preferable to
delete a number of nodes.  In order to eliminate finite size effects, it
is desirable to block all lattices down to the same target number of nodes.
As the volume of the initial lattice is varied, the resulting blocked theories
embody the physics at a range of length scales.  Two operators are used here
to track the renormalization group flows: the volume and a term $M$ defined
by
$$M = \sum_{i \in N_0} \ln \left( {O_i \over {D + 1}} \right) $$
where $O_i$ is the number of simplexes containing the node $i$.
If $M$ is added to the action
$$S_M = \mu M $$
it corresponds to the addition of a measure term [9].

The model of dynamical triangulations considered first has the action
$$ S = \alpha N_0 - \beta N_D$$
where $D$ is either three or four [10].
$N_0$ is the number of nodes with $\alpha$ corresponding to Newton's constant
while $N_D$ is the volume so that $\beta$ is the cosmological constant.
Figure 1 shows a plot of
$<M>$ versus the volume as a function of the degree of blocking
for various values of $\alpha$ in four dimensions.  The renormalization group
flows for the smaller values of $\alpha$ flow toward the right, toward
decreasing node density, which is the expected behavior in the crumpled phase.
For larger $\alpha$ flows are to the left, a different behavior, indicative
of a flow toward a different trivial fixed point, presumably associated with
the smooth phase.  The presence of an intermediate type of flow suggests that
a non-trivial fixed point may be nearby.

\begin{figure}
\centering
\epsfxsize=3.0in \epsfbox{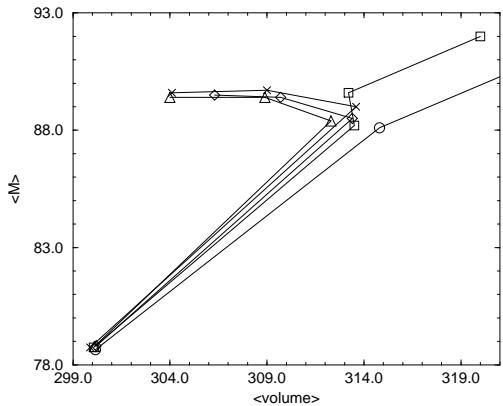}
\caption{Renormalization group flows in the three-dimensional dynamical
         triangulation model near its transition.}
\label{threed}
\end{figure}

In three dimensions, shown in Fig. 2, there are similar flows to the left
and to the right, but no intermediate flow.  This is consistent with the
strongly first order character of the transition.  

\begin{figure}
\centering
\epsfxsize=3.0in \epsfbox{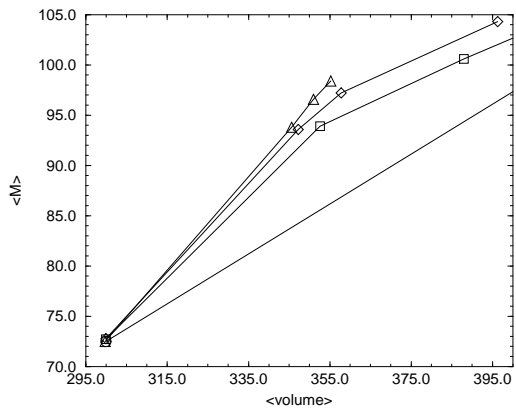}
\caption{Renormalization group flows in the three-dimensional dynamical
         triangulation model with a measure term included.}
\label{wmeas}
\end{figure}

The phase diagram of both the three-dimensional and four-dimensional theories
can be expanded by adding the previously defined measure term to the action.
Recent work indicates that the $\mu = 0$ transitions are a part of a first
order line in the $(\alpha,\mu)$ plane [11].  If this line terminates in a
critical
endpoint, that would provide the second order phase transition required for
taking a continuum limit.  This talk considers whether there is still a
transition at $\mu = -1$ in three dimensions.  Flows for various values of
$\alpha$ are given in Fig. 3.  There is no evidence of a transition.  More
recent work at very large values of $\alpha$ has discovered flows like that
of the smooth phase [12], but that is not necessarily inconsistent with
behavior past the end of a first order line.  More work will be necessary to
establish that the first order line actually ends.


\begin{thebibliography}{99}
\bibitem{rg1}    K. G. Wilson, Rev. Mod. Phys. 47, 773 (1975).

\bibitem{rg2}    K. G. Wilson, in {\it Recent Developments in Gauge Theories},
                 ed. G 't Hooft (Plenum Press, N. Y. 1980).

\bibitem{rg7}    M. E. Fisher, {\it Scaling, Universality and Renormalization
                 Group Theory}, in {\it Critical Phenomena} Lecture Notes in
                 Physics, Vol. 186, ed F. J. W. Hahne (Springer-Verlag, Berlin,
                 1983).

\bibitem{ising}  G. S. Pawley, R. H. Swendsen, D. J. Wallace, and K. G. Wilson,
                 Phys. Rev. B29, 4030 (1984).

\bibitem{renken} R. L. Renken, Phys. Rev. D50, 5130 (1994).

\bibitem{rck}    R. L. Renken, S. M. Catterall, and J. B. Kogut,
                 Phys. Lett. B345, 422 (1995).

\bibitem{thorli} G. Thorleifsson and S. M. Catterall, SU-4240-619.

\bibitem{catter} S. M. Catterall, Computer Physics Comm. 87, 409, 1995.

\bibitem{brugmann} B. Brugmann and E. Marinari, Phys. Rev. Lett. 70,
                   1908, 1993.

\bibitem{dtreview} S. Catterall, ``Lattice Quantum Gravity: Review and Recent
                   Developments," hep-lat/9510008 and D. Johnston, ``Gravity
                   and Random Surfaces," hep-lat/9607021.

\bibitem{ckrnew} S. Catterall, J. Kogut, and R. Renken, manuscript in
                 preparation.

\bibitem{renken2} R. L. Renken, ``A Renormalization Group for Dynamical
                  Triangulations in Arbitrary Dimensions," hep-lat/9607074.

\end{thebibliography}
\end{document}